\documentclass[reprint,aps,prl]{revtex4-1}

\renewcommand{\Re}{\mathop{\mathrm{Re}}}

\usepackage{graphicx}
\usepackage{amsmath, amsthm, amssymb}
\usepackage{epstopdf}
\usepackage{mathtools}
\usepackage{circuitikz}
\usepackage{dcolumn}
\usepackage[tight]{subfigure}
\usepackage[normalem]{ulem}
\usepackage{verbatim}
\usepackage{inputenc}
\usepackage{color}
\usepackage{bm} 
\usepackage{natbib}
\usepackage{xspace}
\usepackage{hyperref} 
\usepackage{mathbbol}

\begin{document}

\title{Nonlinear thermoelectricity with electron-hole symmetric systems}
\author{G. Marchegiani}
\email{giampiero.marchegiani@nano.cnr.it} 

\affiliation{NEST Istituto Nanoscienze-CNR and Scuola Normale Superiore, I-56127 Pisa, Italy}

\author{A. Braggio}
\email{alessandro.braggio@nano.cnr.it} 
\affiliation{NEST Istituto Nanoscienze-CNR and Scuola Normale Superiore, I-56127 Pisa, Italy}

\author{F. Giazotto}
\email{francesco.giazotto@sns.it} 
\affiliation{NEST Istituto Nanoscienze-CNR and Scuola Normale Superiore, I-56127 Pisa, Italy}

\date{\today}

\begin{abstract}
In the linear regime, thermo-electric effects between two conductors are possible only in the presence of an explicit breaking of the electron-hole symmetry. We consider a tunnel junction between two electrodes and show that this condition is no longer required outside the linear regime. In particular, we demonstrate that a thermally-biased junction can display an absolute negative conductance (ANC), and hence thermo-electric power, at a small but finite voltage bias, provided that the density of states of one of the electrodes is gapped and the other is monotonically decreasing. We consider a prototype system that fulfills these requirements, namely a tunnel junction between two different superconductors where the Josephson contribution is suppressed.
We discuss this nonlinear thermo-electric effect based on the spontaneous breaking of electron-hole symmetry in the system, characterize its main figures of merit and discuss some possible applications.
\end{abstract}

\maketitle

\textit{Introduction.} Recently, thermal transport at the nanoscale and the field of quantum thermodynamics have attracted a growing interest~\cite{Benenti2017,DiVentraRMP,KosloffEntropy,SeifertRepProgPhys,Muhonen2012,CampisiRMP,GiazottoRMP,FornieriReview,Brunner2012,SeifertPRL2015,PolettiniPRL,Verley2014,SeifertPRL2018,ManikandanPRL}. In particular, thermo-electric systems have been extensively investigated~\cite{LambertPRB,SeifertPRL2013,Sothmann_2014,Ozaeta2014,Esposito_2009,Vischi2019,WhitneyPRL2014,ThermophasePRL,Ronetti2016,GiazottoTransistor,MarchegianiEngine,Kamp}, since they provide a direct thermal-to-electrical power conversion. In a two-terminal system, a necessary condition for thermoelectricity in the linear regime, i.e., for a small voltage $V$ and a small temperature bias $\Delta T$, is breaking the electron-hole (EH) symmetry which results in the transport property $I(V,\Delta T)\neq-I(-V,\Delta T)$, where $I$ is the charge current flowing through the two-terminal system. In fact, if $I(V,\Delta T)=-I(-V,\Delta T)$, it follows $I(0,\Delta T)=0$, and hence a null thermopower, irrespectively of the temperature bias $\Delta T$. Nonlinear thermoelectric effects have been also investigated~\cite{AzemaPRB,Kim2014,Zimbovskaya,SVILANS201634,SanchezSerra,Boese_2001,SANCHEZ20161060,WhitneyPRBNonlinear,ErdmanPRB}, even in systems where $I(0,\Delta T)=0$~\cite{SanchezLopez}, but the EH symmetry breaking is always assumed. For metals, the EH symmetry is roughly present for Landau-Fermi liquids at small energies, and indeed thermoelectric effects in real metals are typically small, scaling as $T/T_{\rm F}$, where $T_{\rm F}$ is the Fermi temperature. More generally, a nearly perfect EH symmetry characterizes many interacting systems in the quantum regime, such as superconductors~\cite{Tinkham2004,DeGennesbook} or Dirac materials~\cite{DiracReview}. 

Here, we establish a set of sufficient and universal conditions for finite thermoelectric power $\dot{W}=-IV>0$ in systems where EH symmetry holds $I(V,\Delta T)=-I(-V,\Delta T)$. More precisely, we demonstrate that the electron-hole symmetry breaking which leads to thermoelectricity is driven by the nonlinear temperature difference and asymmetry between the two terminals. 

\textit{Model.} We consider a basic example in quantum transport, namely a tunnel junction, which is also experimentally relevant. The system consists of two conducting electrodes (L, R), coupled through a thin insulating barrier, where quantum tunnelling takes place. In this case, the main contribution to transport is typically given by Landau's fermionic excitations, called quasiparticles. For the purpose of our discussion, we assume each electrode in internal thermal equilibrium, namely the quasiparticle distributions read $f_\alpha(E-\mu_\alpha)=\{1+\exp[(E-\mu_\alpha)/(k_{\rm B}T_\alpha)]\}^{-1}$, where $k_{\rm B}$ is the Boltzmann constant and $T_\alpha$, $\mu_\alpha$ (with $\alpha=$L, R) are the temperatures and the chemical potentials of the quasiparticle systems, respectively. The quasiparticle charge and heat current flowing out of the $\alpha$-electrode (with $\bar\alpha={\rm R}$ when $\alpha={\rm L}$ and \textit{vice versa}) read~\cite{Mahan_book,Tinkham2004}
\begin{equation}
\begin{pmatrix}
I_{\alpha}   \\   
\dot Q_{\alpha}
\end{pmatrix}
=\frac{G_{\rm T}}{e^2}\int_{-\infty}^{+\infty}dE 
\begin{pmatrix}
-e \\   
E_\alpha
\end{pmatrix} N_{\alpha}(E_{\alpha})N_{\bar\alpha}(E_{\bar\alpha})F_\alpha(E_\alpha)
\label{eq:IVandQ}
\end{equation}
where $-e$ is the electron charge, $N_\alpha(E)$ is the quasiparticle density of states (DoS), $F_\alpha(E_\alpha)=f_\alpha(E_\alpha)-f_{\bar\alpha}(E_{\bar\alpha})$,  $E_\alpha=E-\mu_\alpha$, and $G_{\rm T}$ is the conductance of the junction if both the electrodes have constant $N_\alpha$. For simplicity, we assumed spin-degeneracy, and an energy and spin independent tunneling in the derivation of Eq.~\ref{eq:IVandQ}. We consider EH symmetric DoSs: $N_{\alpha}(E)=N_{\alpha}(-E)$ and we define $I=I_{\rm L}$~\footnote{$I_{\rm R}=-I_{\rm L}$ due to charge conservation.}. Under a voltage bias $V\neq 0$, the chemical potentials are shifted: $\mu_{\rm L}-\mu_{\rm R}= -eV$. By exploiting the symmetries, one can show that $I(V,T_{\rm L}, T_{\rm R})=-I(-V,T_{\rm L}, T_{\rm R})$ and $\dot Q_{\alpha}(V,T_{\rm L}, T_{\rm R})=\dot Q_{\alpha}(-V,T_{\rm L}, T_{\rm R})$~\cite{SM}. The expressions of Eq.~$\ref{eq:IVandQ}$ respect the thermodynamic laws~\cite{Benenti2017,WhitneyPRBNonlinear,Prigogine_book,Mazur_book}. In particular, the energy conservation in the junction reads $\dot Q_{\rm L}+\dot Q_{\rm R}+IV=0$ (first law), and the entropy production rate 
$\dot S=-\dot Q_{\rm L}/T_{\rm L}-\dot Q_{\rm R}/T_{\rm R}$ is not negative (second law)~\cite{Benenti2017,YamamotoPRE,WhitneyPRBNonlinear}. 
As a consequence, for $T_{\rm L}=T_{\rm R}=T$ it follows  $IV\geq0$. 
Conversely, for $T_{\rm L}\neq T_{\rm R}$, the condition $IV<0$ is possible. For instance, in a thermo-electric generator, the condition $\dot W=-IV>0$ is thermodynamically consistent with the constraint $\dot S\geq0$ if the efficiency of the conversion $\eta=\dot W/\dot Q_{\rm hot}$ is not larger than the Carnot efficiency, $\eta\leq\eta_{\rm C}=1-T_{\rm cold}/T_{\rm hot}$ ($\dot Q_{\rm hot}>0$ is the heat current from the hot lead). 

Consider the charge current $I$ from Eq.~\ref{eq:IVandQ}. Essentially, the condition on the existence of a thermo-electric power $\dot W>0$ can be expressed as the possibility of having an absolute negative conductance (ANC), $I(V)/V<0$, under a thermal bias. Thanks to EH symmetry, we can focus on $V>0$ and ask whether we can have $I(V,T_{\rm L},T_{\rm R})<0$ for $T_{\rm L}\neq T_{\rm R}$. With no loss of generality, we assume here and in the rest of this work $T_{\rm L}\geq T_{\rm R}$, with $\Delta T=T_{\rm L}-T_{\rm R}$. For $N_{\rm L}(E)=N_{\rm R}(E)$, one can prove that $I(V)\geq 0$ for $V>0$, namely two \emph{different} DoSs are necessary for thermoelectricity in the presence of EH symmetry~\cite{SM}. Our goal is to derive sufficient conditions on the two DoSs which guarantee the existence of thermoelectricity.

To this end, it is convenient to measure the energy $E$ with respect to $\mu_{\rm L}$, i.e., we set $\mu_{\rm L}=0$, $\mu_{\rm R}=eV$.  We rewrite, with simple manipulations, the charge current $I$ of Eq.~\ref{eq:IVandQ} as
\begin{align}
& I=\frac{G_{\rm T}}{e}\int_{0}^{\infty}dE N_{\rm L}(E)f_{\rm L}(E)[N_{\rm R}(E_+)-N_{\rm R}(E_-)]+\nonumber\\
&\frac{G_{\rm T}}{e}\int_{0}^{\infty}dE N_{\rm L}(E)[N_{\rm R}(E_-)f_{\rm R}( E_-)-N_{\rm R}(E_+)f_{\rm R}(E_+)] 
\label{eq:IV-mod}
\end{align}
where $E_{\pm}=E\pm eV$. If $N_{\rm L}$ is a gapped function (with gap $\Delta_{\rm L}$), that is $N_{\rm L}\approx0$ for $|E|<\Delta_{\rm L}$, the second term in Eq.~\ref{eq:IV-mod} is negligible when $eV,k_{\rm B}T_{\rm R}\ll\Delta_{\rm L}$, due to the exponential damping of the cold distribution $f_{\rm R}$ above the gap $\Delta_{\rm L}$. Moreover, for $k_{\rm B}T_{\rm L}\sim\Delta_{\rm L}$, the integrand function in the first term of Eq.~\ref{eq:IV-mod} is finite, owing to the presence of the hot distribution $f_{\rm L}$, and negative when $N_{\rm R}(E)$ is a monotonically decreasing function for $E>\Delta_{\rm L}-eV$. In conclusion, even with EH symmetric DoSs, the presence of a gap in the hot electrode DoS and the monotonically decreasing function in the cold electrode DoS may generate an ANC, and hence thermoelectricity  $\dot W=-IV>0$. This is the crucial result of this work, and can be applied in a quite general setting~\cite{SM}. Below, we discuss the
main features of this nonlinear thermoelectric effect for an experimentally suitable EH symmetric system: a tunnel junction between two Bardeen-Cooper-Schrieffer (BCS~\cite{BCS}) superconductors (SIS junction). 

\begin{figure}[tp]
	\begin{centering}
		\includegraphics[width=0.45\textwidth]{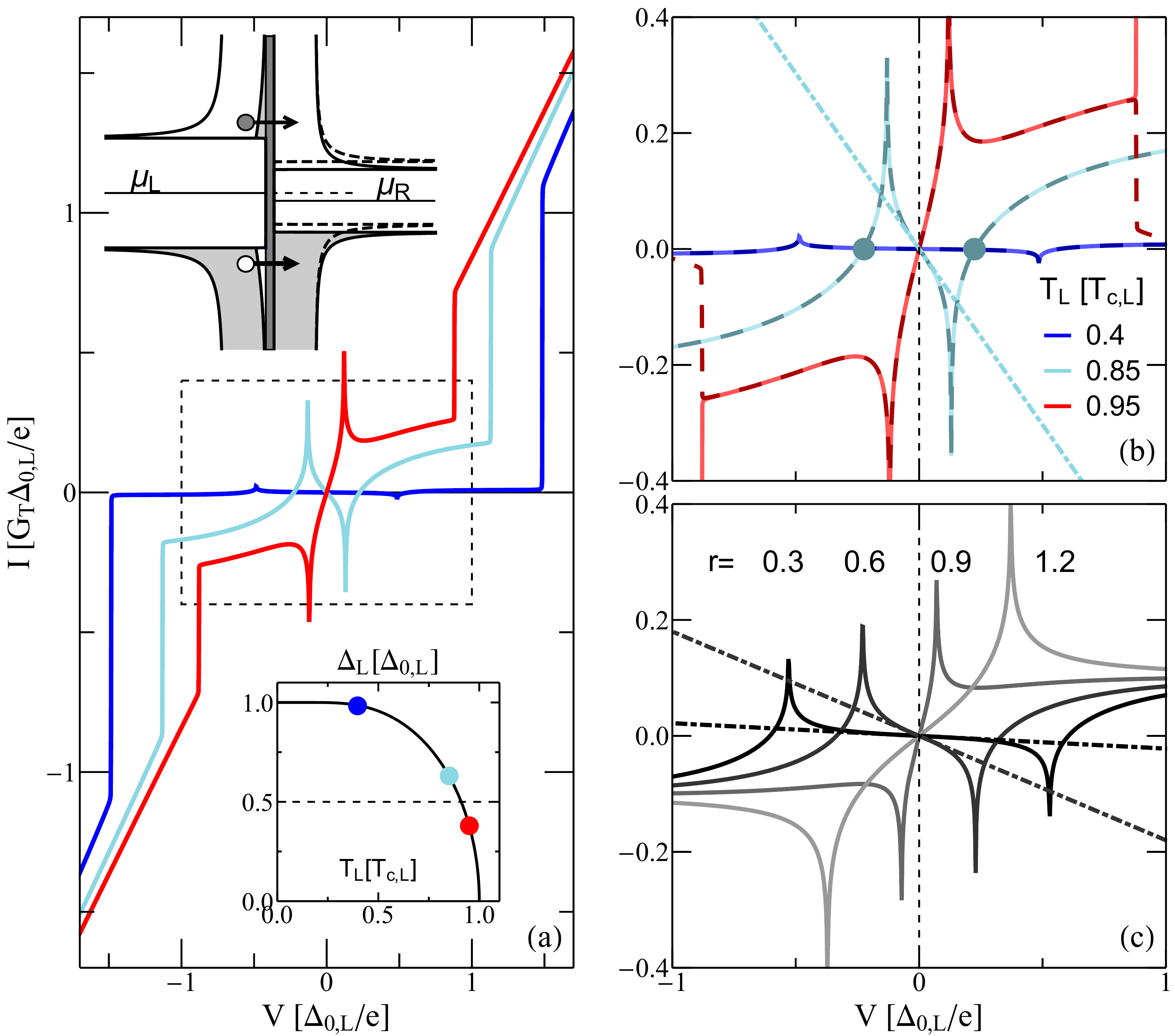}
		\caption{(color online). \textbf{(a)} Quasiparticle current-voltage characteristic of a thermally biased tunnel junction between two superconductors (SIS junction) for $T_{\rm R}=0.01 T_{\rm c,L}$, $r=0.5$ and different values of $T_{\rm L}>T_{\rm R}$. The curves display absolute negative conductance (ANC) and thermoelectric power $\dot W=-IV>0$ at small voltage bias if $\Delta_{\rm L}(T_{\rm L})>\Delta_{\rm 0,R}$.  Top inset: energy band diagram of the SIS junction. The combination of the gap in the hot electrode (left) and the monotonically decreasing DoS above gap of the cold electrode (right) produces a particle current which flows in the opposite direction of the chemical potential gradient. Bottom inset: temperature dependence of the superconducting gap $\Delta_{\rm L}$. Colored points mark the values of $\Delta_{\rm L}(T_{\rm L})$ for the curves displayed in the panel a). The horizontal dashed line intercepts the $\Delta_{\rm L}$ curve at the point where $\Delta_{\rm L}(T_{\rm L})=\Delta_{\rm 0,R}$, i.e., the maximum temperature for the existence of the ANC in panels (a) and (b).  \textbf{b)} Enlargement of the subgap transport in panel (a) (dashed rectangle). Dashed curves give the first term of Eq.~\ref{eq:IV-mod}. The light-blue dots give the values of the Seebeck voltage $V_{\rm S}$. \textbf{(c)} Subgap IV characteristics for  $T_{\rm L}=0.7 T_{\rm c,L}$, $T_{\rm R}=0.01 T_{\rm c,L}$, and different values of $r$. The slopes of the dash-dotted lines in panels (b) and (c) give the values of the ANC at $V\approx0$, as expressed by Eq.~\ref{eq:G0}.
		}
		\label{Fig1}
	\end{centering}
\end{figure}

\emph{SIS junction.} For simplicity, we focus on quasiparticle transport and assume to completely suppress the Josephson contribution occurring in SIS junctions~\cite{FornieriReview,GlazmanPRB,GuarcelloPRApplied}. This condition can be achieved either by considering a junction with a strongly oxidized barrier or by appropiately applying an external in-plane magnetic field. The quasiparticle DoS reads
$N_{\alpha}=\theta(|E|-\Delta_\alpha)|E|/\sqrt{E^2-\Delta_\alpha^2}$~\footnote{The subgap transport in realistic junctions is accounted for with a small parameter  $\Gamma_\alpha$~\cite{Dynes1984}. The DoS reads: $N_{\rm \alpha}=|\Re[(E+i\Gamma_\alpha)/\sqrt{(E+i\Gamma_\alpha)^2-\Delta_\alpha^2}]|$. In the calculations we set $\Gamma_\alpha=10^{-4}\Delta_{0,\alpha}$.},
where $\Delta_{\rm \alpha}(T_{\alpha})$ are the temperature dependent superconducting order parameters. 
In particular, $\Delta_{\rm \alpha}(T_{\rm \alpha}=0)=\Delta_{\rm 0,\alpha}$ and it decreases monotonically with $T_{\rm \alpha}$, following a universal relation, obtained through a self-consistent calculation~\cite{Tinkham2004} (see the bottom inset of Fig.\ref{Fig1}a). It becomes zero when the temperature approaches the critical value $T_{\rm c,\alpha}=\Delta_{\rm 0, \alpha}/(1.764 k_{\rm B})$. We stress that the temperature dependence of $\Delta_{\rm \alpha}$ is not necessary for the mechanism, and it is characteristic of the specific system here considered.

Since $N_{\rm L}(E)\neq N_{\rm R}(E)$ is a necessary condition for thermoelectricity, hereafter we consider the case where the two gaps at zero-temperature differ introducing a parameter $r=\Delta_{\rm 0,R}/\Delta_{\rm 0,L}=T_{\rm c,R}/T_{\rm c,L}$. 

Consider now Eq.~\ref{eq:IV-mod} for a SIS junction. As discussed above, for $eV,k_{\rm B}T_{\rm R}\ll \Delta_{\rm L}(T_{\rm L})$ the second term is negligible and $I(V)$ is given entirely by the first contribution. To have ANC, two conditions must apply: i) the hot temperature $T_{\rm L}$ must be of the order of the gap, $k_{\rm B}T_{\rm L}\lesssim \Delta_{\rm L}(T_{\rm L})$, due to the presence of $f_{\rm L}(E)$ (but necessarily smaller than $T_{\rm c,L}$ for the superconductivity to survive), ii) the term in the square bracket must be negative.
Since the BCS DoS $N_{\rm R}(E)$ is monotonically decreasing only for $E>\Delta_{\rm R}(T_{\rm R})$, the two conditions require $\Delta_{\rm L}(T_{\rm L})-\Delta_{\rm R}(T_{\rm R})> 0$. Being $\Delta_{\rm L}(T_{\rm L})$ a monotonically decreasing function, the conditions are met only if the hot superconductor has the larger gap. Thus, a necessary condition for ANC is $r<1$ when $T_{\rm L}>T_{\rm R}$. Conversely, by inverting the temperature gradient, i e., $T_{\rm R}>T_{\rm L}$, the thermoelectricity requires $\Delta_{\rm R}(T_{\rm R})-\Delta_{\rm L}(T_{\rm L})> 0$ and the proper conditions are met for $r>1$. The origin of the thermoelectricity can be intuitively understood in the energy band diagram in the top inset of Fig.~\ref{Fig1}a, drawn for $T_{\rm L}>T_{\rm R}$ and $\mu_{\rm L}>\mu_{\rm R}$. The net current is given by the difference of the particle (fill circle) and the holes (empty circle) contributions. They exactly cancel out at $V=0$, due to EH symmetry. For $V\neq 0$, the shifting of $N_{\rm R}$ decreases(increases) the particles(holes) contribution, due to locally monotonic decreasing behavior. As a consequence, the particle current flows in the opposite direction of the chemical potential gradient.

 Figure~\ref{Fig1}a displays the IV characteristics for $r=0.5$, $T_{\rm R}=0.01 T_{\rm c,L}$ and different values of $T_{\rm L}> T_{\rm R}$. The evolution is linear $I\simeq G_{\rm T}V$ at large bias $eV> \Delta_{\rm L}(T_{\rm L})+\Delta_{\rm R}(T_{\rm R})$ and strongly nonlinear within the gap, i.e., for $eV<\Delta_{\rm L}(T_{\rm L})+\Delta_{\rm R}(T_{\rm R})$. Figure~\ref{Fig1}b gives an enlarged view of the subgap transport displayed in Fig.~\ref{Fig1}a (dashed rectangle). Within the gap, the curves display characteristic peaks at $eV_{\rm peak}=\pm|\Delta_{\rm L}(T_{\rm L})-\Delta_{\rm R}(T_{\rm R})|\sim \pm|\Delta_{\rm L}(T_{\rm L})-\Delta_{\rm 0,R}|$, due to the matching of the BCS singularities in the DoSs. Interestingly, the curves display a significant ANC, and hence thermoelectricity, for intermediate values of $T_{\rm L}$. Furthermore, the thermoelectric effect is negligible if $\Delta T=T_{\rm L}- T_{\rm R}$ is too low and it is absent when $\Delta_{\rm L}(T_{\rm L})<\Delta_{\rm R}(T_{\rm R})$. The contributions due to the first term of Eq.~\ref{eq:IV-mod} are displayed with dashed lines in Fig.~\ref{Fig1}b. As argued above, they yield a good approximation for $eV<\Delta_{\rm L}$. The dependence of the IV characteristics on $r$ is visualized in Fig.~\ref{Fig1}c for $T_{\rm L}=0.7T_{\rm c,L}>T_{\rm R}=0.01T_{\rm c,L}$. In particular, the ANC is present only when $\Delta_{\rm 0,R}<\Delta_{\rm L}(T_{\rm L})\sim0.83\Delta_{\rm 0,L}$, namely for $r\lesssim0.83$.
 
 For $V\sim 0$, the IV characteristic is approximately linear and, by using the first term of Eq.~\ref{eq:IV-mod}, we can derive an expression for the negative conductance~\cite{SM}, namely
\begin{equation}
G_0=\lim_{V\rightarrow 0}\frac{I(V)}{V}=
-2G_{\rm T}\Delta_{\rm 0,R}^2\int_{\Delta_{\rm L}(T_{\rm L})}^{\infty}dE \frac{N_{\rm L}(E)f_{\rm L}(E)}{(E^2-\Delta_{\rm 0,R}^2)^{3/2}},
\label{eq:G0}
\end{equation} 
valid for $T_{\rm R}\ll T_{\rm c,R}$ and $\Delta_{\rm L}(T_{\rm L})>\Delta_{\rm 0,R}$. This negative slope is shown in Fig.~\ref{Fig1}b-c for some curves with dotted-dashed lines, which perfectly represent the linear-in-bias behaviour. 

We stress that the existence of the ANC in a thermally biased SIS junction is not discussed in the literature to the best of our knowledge. This is not totally surprising, since the ANC can be observed only for $r\neq 1$ and higher temperature of the larger gap superconducting electrode $T_{\rm L}\lesssim T_{\rm c,L}$. This effect is reminiscent of the ANC predicted~\cite{Aronov1975} and observed in experiments on nonequilibrium superconductivity, with particles injection~\cite{Gershenzon1986,Gershenzon1988,Gijsbertsen1996} or microwave irradiation~\cite{NagelPRL}.

\emph{Thermoelectric figures of merit.}
Due to the nonlinear nature of the effect, we cannot rely on the standard figures of merit for linear thermo-electric effects. Yet, in the nonlinear regime we can still define the Seebeck voltage $V_{\rm S}$ which corresponds to the voltage developed by the thermal bias $\Delta T$ at open circuit.
 Consider, for instance, the light-blue curve in Fig.~\ref{Fig1}b, where there is thermoelectricity $\dot W>0$. Clearly, the curve crosses the x-axis in $V=0$, as required by EH symmetry. Furthermore, if there is ANC at low voltage ($I/V<0$) and an Ohmic behaviour at large voltage ($I/V\sim G_{\rm T}>0$), there will be, at least, two finite values $V=\pm V_{\rm S}\neq0$ where $I(V)=0$ (see marked points in Fig.~\ref{Fig1}b).
\begin{figure}[tp]
	\begin{centering}
		\includegraphics[width=0.45\textwidth]{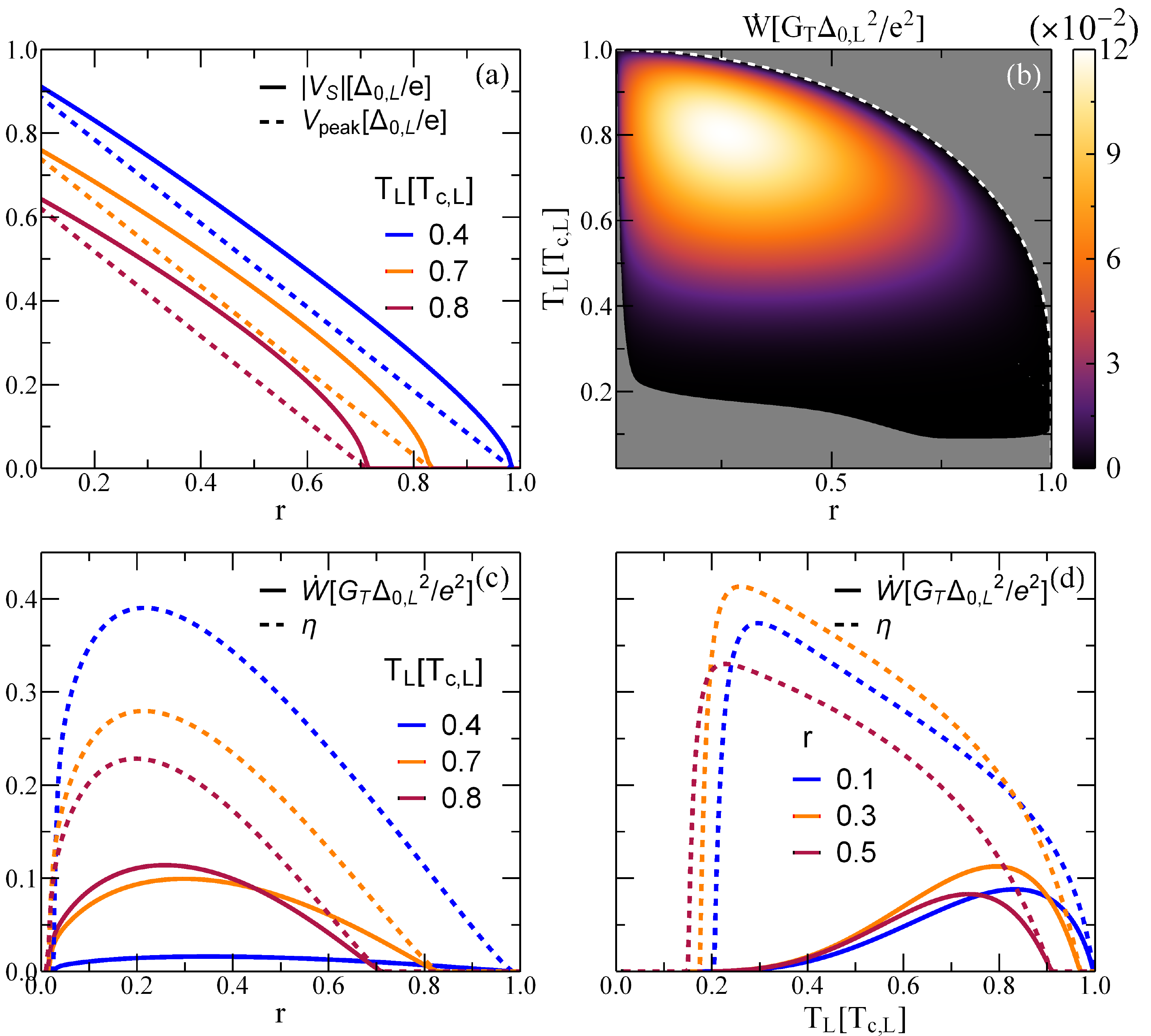}
		\caption{(color online). Thermoelectric figures of merit for a SIS junction.  \textbf{(a)} Seebeck voltage vs $r$ for $T_{\rm R}=0.01T_{\rm c,R}$ and some values of $T_{\rm c,L}$ (solid). The voltage corresponding to the singularity matching peak is displayed for a comparison (dashed).  \textbf{(b)} Density plot of the thermoelectric power $\dot W=-IV$ vs $r$ and $T_{\rm L}$ for $T_{\rm R}=0.01T_{\rm c,L}$. In the gray region the thermoelectric effect is absent, i.e., the junction is dissipative $\dot W<0$. The white dashed curve displays the equation $\Delta_{\rm L}(T_{\rm L})=\Delta_{\rm 0,R}$.  \textbf{(c),(d)} Cuts of Fig.~\ref{Fig2}b for particular values of $T_{\rm L}$ and $r$, respectively. The correspondent thermoelectric efficiency $\eta=\dot W/\dot Q_{\rm L}$ is plotted with dashed lines.
		}
		\label{Fig2}
	\end{centering}
\end{figure}

Figure~\ref{Fig2}a displays $|V_{\rm S}|$ as a function of $r$ for $T_{\rm R}=0.01 T_{\rm c,R}$ and some values of $T_{\rm L}>T_{\rm R}$ (solid lines). The curves show some characteristic features: i) for a given $T_{\rm L}$, $|V_{\rm S}|$ decreases monotonically with $r$ and it is zero when $r$ is larger than some critical value (depending on $T_{\rm L}$), ii) for a given $r$, $|V_{\rm S}|$ \textit{decreases} when the temperature $T_{\rm L}$, that is proportional to the temperature difference $\Delta T$, is increased, something that differ with the usual linear thermoelectricity. These features can be qualitatively understood by comparing $V_{\rm S}$ with the matching peak value $V_{\rm peak}=[\Delta_{\rm L}(T_{\rm L})-\Delta_{\rm R}(T_{\rm R})]/e$ (dashed curves in Fig.~\ref{Fig2}a). In fact, the magnitude of $V_{\rm S}$ is correlated to $V_{\rm peak}$, i.e., $|V_{\rm S}|\geq V_{\rm peak}$ when there is thermoelectricity (see Fig.~\ref{Fig1}b,c). By definition, for a given $T_{\rm L}$, $V_{\rm peak}$ decreases almost linearly with $r$, i.e., $eV_{\rm peak}/\Delta_{\rm 0,L}\sim\Delta_{\rm L}(T_{\rm L})/\Delta_{\rm 0,L}-r$. This explains also the temperature evolution, since $\Delta_{\rm L}(T_{\rm L})$ is a monotonically decreasing function. In particular, when $r$ is larger than a critical value depending on $T_{\rm L}$, i.e., $r\gtrsim\Delta_{\rm 0,R}/\Delta_{\rm L}(T_{\rm L})$, $V_{\rm S}$ goes to zero since $\Delta_{\rm L}(T_{\rm L})<\Delta_{\rm R}(T_{\rm R})$, i.e., there is no thermoelectricity. For $r=0.3$, an effective nonlinear Seebeck coefficient $\mathcal S=V_{\rm S}/\Delta T$ can reach values as large as  $\sim 0.8\Delta_{\rm 0,L}/(0.4 e T_{\rm c,L})=2\times1.764 k_{\rm B}/e\sim 300 \mu$V/K.

Now, we consider the thermo-electric power $\dot W=-IV$. For simplicity, we evaluate it at $V_{\rm peak}$, where it is approximately maximum~\cite{SM}, namely $-I(V_{\rm peak})V_{\rm peak}\sim \max_{V}(-IV)$. Figure~\ref{Fig2}b displays the density plot of $\dot W$ as a function of $r$ and $T_{\rm L}$ for $T_{\rm R}=0.01T_{\rm c,L}$. The thermo-electric power is absent if $T_{\rm L}\leq 0.1T_{\rm c,L}$, irrespectively of $r$. Furthermore, it is zero when $\Delta_{\rm L}(T_{\rm L})<\Delta_{\rm 0,R}$ (the dashed white line in Fig.~\ref{Fig2}b displays the curve $\Delta_{\rm L}(T_{\rm L})=\Delta_{\rm 0,R}$). The maximum value of $\dot W$ is obtained at $r\sim 0.25$ and $T_{\rm L}=0.8T_{\rm c,L}$ and it yields $\dot W_{\rm max}\sim 0.11~G_{\rm T}\Delta_{\rm 0,L}^2/e^2$. For an aluminum based ($\Delta_{\rm 0,L}/e\sim200\mu$V)  tunnel junction with $G_{\rm T}=(1 {\rm k}\Omega)^{-1}$, the maximum is $\dot W_{\rm max}\sim 4$ pW.

For a better characterization, we consider cuts of Fig.~\ref{Fig2}b for specific values of $T_{\rm L}$ (solid curves in Fig.~\ref{Fig2}c) and $r$ (solid curves in Fig.~\ref{Fig2}d). In both the panels, we add the corresponding thermoelectric efficiency  $\eta=\dot W/\dot Q_{\rm L}$ (dashed curves). Interestingly, the highest absolute efficiency with respect to $r$ is obtained almost in correspondence of the maximum power $\eta_{\rm max}\sim 0.4$ (see Fig.~\ref{Fig2}c). Conversely, the best condition for $\eta$ as a function of $T_{\rm L}$ does not coincide with the condition for maximum power (see Fig.~\ref{Fig2}d), although $\eta$ is quite high even at the best condition in terms of power $\eta_{\dot W_{\rm max}}=\eta(T_{\rm L}=0.8T_{\rm c,L})\sim 0.22$ (orange line in Fig.~\ref{Fig2}d).

\emph{Spontaneous symmetry breaking.} 
\begin{figure}[tp]
	\begin{centering}
		\includegraphics[width=0.45\textwidth]{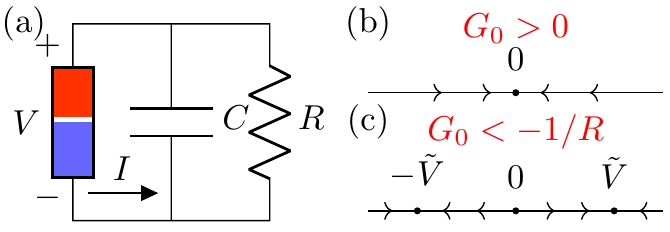}
		\caption{(color online). \textbf{(a)} Circuital scheme: the junction is a nonlinear element with characteristic $I(V,T_{\rm L},T_{\rm R})$ and capacitance $C$, connected to a generic load R. \textbf{(b)-(c)} Phase portrait for the voltage dynamics across the system. (b) In the absence of thermoelectricity, $G_0>0$ and the voltage relaxes to 0, due to the dissipation in the load. (c) In the presence of thermoelectricity and for $G_0<-1/R$, the zero-voltage solution is unstable and a voltage, either $\pm\tilde V$, spontaneously develops across the junction.  
		}
		\label{Fig3}
	\end{centering}
\end{figure}
Here we discuss the experimental consequences of thermoelectricity in terms on the junction's dynamics. We consider a minimal circuital setup, displayed in Fig.~\ref{Fig3}a. The junction is modeled as a nonlinear element of characteristic $I(V,T_{\rm L},T_{\rm R})$ and capacitance $C$, in parallel with a load external circuit of resistance $R$. The evolution is obtained by requiring the current conservation in the circuit,
\begin{equation}
I(V,T_{\rm L},T_{\rm R})=-C \dot V-\frac{V}{R},
\label{sys:circuit}
\end{equation}
where the dot denotes the time ($t$) derivative. The stationary points are obtained by setting $\dot V=0$ in Eq.~\ref{sys:circuit} and read $V(t)=\tilde V$, where $\tilde V$ is a solution of the implicit equation $RI(\tilde V,T_{\rm L},T_{\rm R})+\tilde V=0$. Since $I(V,T_{\rm L},T_{\rm R})=-I(-V,T_{\rm L},T_{\rm R})$, the equation has an odd number of solutions and $\tilde V=0$ is always a solution, irrespectively of $R,T_{\rm L},T_{\rm R}$. The stability of these solutions can be acquired by linearizing Eq.~\ref{sys:circuit}, namely $ \dot v=-C^{-1}[G(\tilde V)+1/R]v$, where $v=V-\tilde V$ and $G(\tilde V)=dI/dV|_{V=\tilde V}$. The solution is stable if the term in the square bracket is positive and unstable otherwise.

In the absence of thermoelectricity, $IV\geq 0$ and the zero-bias conductance of the junction is positive $G_0\geq 0$. Thus, $\tilde V=0$ is the unique solution of Eq.~\ref{sys:circuit} and it is stable (see Fig.~\ref{Fig3}b). Conversely, when we apply a temperature gradient and the SIS junction displays thermoelectricity, $G_0<0$ (see Eq.~\ref{eq:G0}), and additional solutions at finite voltages are possible. In particular, for sufficiently large values of the load, such as $R>-G_0^{-1}$ there are three solutions $V=0,\pm \tilde V$, and $G(\pm\tilde V)>0$. As a consequence, any voltage signal across the device evolves toward one of the two values $\pm \tilde V$, depending on the initial conditions (see Fig.~\ref{Fig3}c). Namely, the combination of a sufficiently strong thermal gradient and the voltage polarization imposed by the external circuit leads to a \textit{spontaneous breaking} of EH symmetry. Moreover, the bi-stability of the stationary voltage may be used to design a volatile thermo-electric memory or a switch~\cite{SM}. In a more general setting which includes inductive effects, the instability of the zero-voltage state can generate also a self-sustained oscillatory dynamics~\cite{SM,BenentiPRE,ALICKI201771}.

\textit{Conclusions.} In summary, we discussed a general thermo-electric effect occurring in systems with EH symmetry in the nonlinear regime. For a two-terminals tunneling system, two sufficient conditions are required for thermoelectricity: i) the hot electrode has a gapped DoS, ii) the cold electrode has a locally monotonically decreasing DoS. In particular, we investigated a prototype system: a tunnel junctions between two different BCS superconductors. We displayed the relevant figures of merit and showed that a thermoelectric voltage spontaneously develops across the system, under proper conditions. Our results may be extended to different classes of materials, including hybrid ferromagnetic-superconducting junctions or low-dimensional quantum systems (dots or wires). This work can represent a promising step in the exploration of thermo-electric effects in the nonlinear regime.  
 
\begin{acknowledgments}
We thank Robert Whitney, David S\'anchez, Tom\'a\v{s} Novotn\'y, Bj\"orn Sothmann, and Giuliano Benenti for discussions and comments. We acknowledge the Horizon research and innovation programme under grant agreement No. 800923 (SUPERTED) for partial financial support. A.B. acknowledges the CNR-CONICET cooperation program Energy conversion in quantum nanoscale hybrid devices, the Royal Society through the International Exchanges between the UK and Italy (Grant No. IES R3 170054 and IEC R2 192166) and the SNS-WIS joint lab QUANTRA.
\end{acknowledgments}
\bibliographystyle{apsrev4-2}

\end{document}


\title{Supplemental Material:  Nonlinear thermoelectricity with electron-hole symmetric systems}
\author{G. Marchegiani}
\email{giampiero.marchegiani@nano.cnr.it} 

\affiliation{NEST Istituto Nanoscienze-CNR and Scuola Normale Superiore, I-56127 Pisa, Italy}

\author{A. Braggio}
\email{alessandro.braggio@nano.cnr.it} 
\affiliation{NEST Istituto Nanoscienze-CNR and Scuola Normale Superiore, I-56127 Pisa, Italy}

\author{F. Giazotto}
\email{francesco.giazotto@sns.it} 
\affiliation{NEST Istituto Nanoscienze-CNR and Scuola Normale Superiore, I-56127 Pisa, Italy}

\date{\today}

\maketitle
\section{Symmetries of the charge and the heat currents}
The symmetries of the charge and the heat current in the presence of electron-hole symmetry of the two DoSs, i.e., for $N_\alpha(E)=N_\alpha(-E)$ with $\alpha$=L,R, are better discussed by measuring the energy with respect to the mean value between the chemical potentials $\mu_0=(\mu_{\rm L}+\mu_{\rm R})/2$. Namely, one sets $\mu_{\rm L}=-eV/2$ and $\mu_{\rm R}=eV/2$.
\subsection{Charge current}
The charge current reads
\begin{equation}
I=\frac{G_{\rm T}}{e}\int_{-\infty}^{+\infty}dE 
N_{\rm L}\left(E+\frac{eV}{2}\right)
N_{\rm R}\left(E-\frac{eV}{2}\right)
\left[f_{\rm R}\left(E-\frac{eV}{2}\right)-f_{\rm L}\left(E+\frac{eV}{2}\right)\right].
\end{equation}
It is convenient to define $\tilde V=eV/2$ and write the integral dividing the positive from the negative energy states (which for $V\to 0$ coincide exactly with the particle and the hole contributions, respectively):
\begin{align}
I&=
\frac{G_{\rm T}}{e}\int_{0}^{+\infty}dE 
N_{\rm L}(E+\tilde V)
N_{\rm R}(E-\tilde V)
\left[f_{\rm R}(E-\tilde V)-f_{\rm L}(E+\tilde V)
\right]
\nonumber\\
&+\frac{G_{\rm T}}{e}\int_{-\infty}^{0}dE 
N_{\rm L}(E+\tilde V)
N_{\rm R}(E-\tilde V)
\left[
f_{\rm R}(E-\tilde V)
-f_{\rm L}(E+\tilde V)
\right].
\end{align}
Upon replacing $E\rightarrow -E$ in the second term, and by using the EH symmetry and the identity $f_\alpha(-E)=1-f_\alpha(E)$, one obtains
\begin{align}
I&=\frac{G_{\rm T}}{e}\int_{0}^{+\infty}dE 
N_{\rm L}(E+\tilde V)
N_{\rm R}(E-\tilde V)
\left[f_{\rm R}(E-\tilde V)
-f_{\rm L}(E+\tilde V)
\right]
\nonumber\\
&-\frac{G_{\rm T}}{e}\int_{0}^{+\infty}dE 
N_{\rm L}(E-\tilde V)
N_{\rm R}(E+\tilde V)
\left[f_{\rm R}(E+\tilde V)
-f_{\rm L}(E-\tilde V)
\right],
\end{align}
which is manifestly odd in the voltage bias $V$. As a consequence, the thermoelectric power $\dot W=-IV$ is even in the voltage bias $V$.

Moreover, this last expression can be used to prove that the condition $IV<0$ is impossible for $N_{\rm L}=N_{\rm R}=N$ irrespectively of the temperatures $T_{\rm L},\ T_{\rm R}$. In fact, by collecting with respect to the Fermi functions of the two leads, one finds
\begin{align}
I&=\frac{G_{\rm T}}{e}\int_{0}^{+\infty}dE 
N(E+\tilde V)
N(E-\tilde V)
\left[f_{\rm L}(E-\tilde V)
-f_{\rm L}(E+\tilde V)
\right]
\nonumber\\
&+\frac{G_{\rm T}}{e}\int_{0}^{\infty}dE 
N(E+\tilde V)
N(E-\tilde V)
\left[f_{\rm R}(E-\tilde V)
-f_{\rm R}(E+\tilde V)\right].
\end{align}
Due to EH symmetry, we can focus on $V>0$. Since the Fermi distribution is monotonically decreasing, the integrand in both the contributions is not negative for any bias $V$ and any energy $E$. Thus, the current is always not negative. Therefore, the response of the system is purely dissipative when $N_{\rm L}=N_{\rm R}$ and no thermoelectricity is possible.
\subsection{Heat current}
The heat current from the left electrode read
\begin{equation}
\dot Q_{\rm L}=
\frac{G_{\rm T}}{e^2}\int_{-\infty}^{+\infty}dE 
(E+\tilde V)
N_{\rm L}(E+\tilde V)
N_{\rm R}(E-\tilde V)
\left[
f_{\rm L}(E+\tilde V)-
f_{\rm R}(E-\tilde V)
\right].
\end{equation}
We can perform the same transformations of the previous section and write down
\begin{align}
\dot Q_{\rm L}&=
\frac{G_{\rm T}}{e^2}\int_{0}^{+\infty}dE 
(E+\tilde V)
N_{\rm L}(E+\tilde V)
N_{\rm R}(E-\tilde V)
\left[
f_{\rm L}(E+\tilde V)-
f_{\rm R}(E-\tilde V)
\right]
\nonumber\\
&+\frac{G_{\rm T}}{e^2}
\int_{0}^{+\infty}dE 
(E-\tilde V)
N_{\rm L}(E-\tilde V)
N_{\rm R}(E+\tilde V)
\left[
f_{\rm L}(E-\tilde V)-
f_{\rm R}(E+\tilde V)
\right],
\end{align}
which is manifestly even in the voltage bias $V$. Symilar considerations can be made for $\dot Q_{\rm R}$.
\section{Features of nonlinear thermoelectricity in gapped systems}
We start from the charge current expression given in Eq.~1 of the main text:
\begin{equation}
I= \frac{G_{\rm T}}{e}\int_{-\infty}^{+\infty}dE N_{\rm L}(E)
N_{\rm R}(E-eV)[f_{\rm R}(E-eV)-f_{\rm L}(E)].
\label{eq:IV}
\end{equation}
We can rewrite this expression in an alternative way, upon summing and subtracting $f_{\rm R}(E)$ in the square bracket and by using the EH symmetry of the DoS.
\begin{align}
I= &\frac{G_{\rm T}}{e}\int_{-\infty}^{+\infty}dE N_{\rm L}(E)
N_{\rm R}(E-eV)[f_{\rm R}(E-eV)-f_{\rm R}(E)]\nonumber\\
+&\frac{G_{\rm T}}{e}\int_{0}^{+\infty}dE N_{\rm L}(E)[N_{\rm R}(E+eV)-N_{\rm R}(E-eV)][f_{\rm L}(E)-f_{\rm R}(E)].
\label{eq:altIV}
\end{align}
The first term is equal to the current through the junction when the two electrodes have temperature $T_{\rm R}$ and it is always larger than zero. The second term may be negative and it is responsible of the thermoelectric contribution.
In fact, if we consider a gapped DoS on the left side $N_{\rm L}(E)=\hat N_{\rm L}(E)\theta(|E|-\Delta_{\rm L})$ and we focus on $eV<\Delta_{\rm L}$ and $k_{\rm B}T_{\rm R}\ll \Delta_{\rm L}$, the first integral is exponentially suppressed, and we can replace $f_{\rm R}\rightarrow 0$ in the second integral, recovering the main contribution of Eq.~2 of the main text.
This expression can be used to obtain the $V\rightarrow 0$ limit in a more general fashion
\begin{align}
G_0= 2G_{\rm T}\int_{\Delta_{\rm L}}^{+\infty}dE N_{\rm L}(E)f_{\rm L}(E)\frac {dN_{\rm R}(E)}{dE}.
\label{eq:altG0}
\end{align}
which is negative for a monotonically decreasing DoS in the right electrode. For a SIS junction, we have $dN_{\rm R}/dE=-\Delta_{\rm R}^2(T_{\rm R})/[E^2-\Delta_{\rm R}^2(T_{\rm R})]^{3/2}$ provided that $\Delta_{\rm L}(T_{\rm L})>\Delta_{\rm R}(T_{\rm R})$. In the limit $T_{\rm R}\rightarrow 0$, one recovers Eq.3 of the main text, upon replacing $\Delta_{\rm R}(T_{\rm R})\rightarrow \Delta_{\rm 0,R}$.
\subsection{Toy model I: one delta DoS vs generic EH symmetric DoS}
Here, we wish to discuss additional models, different from the BCS DoS chosen in the main text, where thermoelectricity can be generated, in order to show the generality of the mechanism.
First, we consider a particular gapped electron-hole symmetric DoS $N_{\rm L}(E)=\epsilon_{\rm L}\delta(|E|-\Delta_{\rm L})$ (where $\epsilon_{\rm L}$ is a constant with energy dimension to ensure a dimensionless normalized DoS) and on the other side a generic electron-hole symmetric DoS $N_{\rm R}(E)=N_{\rm R}(-E)$. For such case, one can easily compute the analytical expression of the current
\begin{align}
I=&\frac{G_{\rm T}\epsilon_{\rm L}}{e}\left\{N_{\rm R}(\Delta_{\rm L}-eV)[f_{\rm R}(\Delta_{\rm L}-eV)-f_{\rm L}(\Delta_{\rm L})]+N_{\rm R}(-\Delta_{\rm L}-eV)[f_{\rm R}(-\Delta_{\rm L}-eV)-f_{\rm L}(-\Delta_{\rm L})]\right\}
\nonumber\\
=&\frac{G_{\rm T}\epsilon_{\rm L}}{e}\left\{
N_{\rm R}(\Delta_{\rm L}-eV)[f_{\rm R}(\Delta_{\rm L}-eV)-f_{\rm L}(\Delta_{\rm L})]-N_{\rm R}(\Delta_{\rm L}+eV)[f_{\rm R}(\Delta_{\rm L}+eV)-f_{\rm L}(\Delta_{\rm L})]\right\}
\end{align}
where we used the electron-hole symmetry. Interestingly, this current may become negative for $V>0$, i.e, the thermoelectricity can be generated in this simplified model. In order to better identify the conditions for it, it is convenient to compute the low-bias behavior.
For $V\to0$, we obtain
\begin{align}
\lim_{V\to 0}I&=\frac{2G_{\rm T}\epsilon_{\rm L}}{e}V
\left\{f_{\rm L}(\Delta_{\rm L})
\frac{dN_{\rm R}(E)}{dE}\biggr\rvert_{\Delta_{\rm L}}
-\frac{d[N_{\rm R}(E)f_{\rm R}(E)]}{dE}
\biggr\rvert_{\Delta_{\rm L}}
\right\}
\nonumber\\&=
\frac{2G_{\rm T}\epsilon_{\rm L}}{e}V
\left\{[f_{\rm L}(\Delta_{\rm L})-f_{\rm R}(\Delta_{\rm L})]\frac{dN_{\rm R}(E)}{dE}\biggr\rvert_{\Delta_{\rm L}}
-N_{\rm R}(E)\frac{df_{\rm R}(E)]}{dE}\biggr\rvert_{\Delta_{\rm L}}
\right\}
\end{align}
The thermoelectricity appears when the current counterflows being $I(V)<0$ for $V>0$. 

It is interesting to compute the limit $T_{\rm R}\to0$, where $f_{\rm R}(E)=0$ for a bias smaller than the gap $0<eV<\Delta_{\rm L}$, getting
\begin{align}
I(V,T_{\rm R}\to0)=&\frac{G_{\rm T}\epsilon_{\rm L}}{e}f_{\rm L}(\Delta_{\rm L})\left[
N_{\rm R}(\Delta_{\rm L}+eV)-N_{\rm R}(\Delta_{\rm L}-eV)\right],\nonumber\\
\lim_{V\to 0}\frac{I(V,T_{\rm R}\to0)}{V}=&\frac{2G_{\rm T}\epsilon_{\rm L}}{e}
f_{\rm L}(\Delta_{\rm L})\frac{dN_{\rm R}(E)}{dE}
\biggr\rvert_{\Delta_{\rm L}}
\end{align}
This expression clearly shows that the thermoelectricity is present for monotonously decreasing DoS and, at the linear order in $V$, is proportional to $\partial_{E} N_{\rm R}(E)|_{\Delta_{\rm L}}<0$. Since the Dirac delta DoS corresponds to the gapped DoS for the hot lead and $N_{\rm R}(E)$ is monotonically decreasing, this is the mechanism described in the main text. 

Similarly one can compute the limit for $T_{\rm L}\to 0$ finding
\begin{align}
I(V,T_{\rm L}\to0)=&\frac{G_{\rm T}\epsilon_{\rm L}}{e}f_{\rm L}(\Delta_{\rm L})\left[
N_{\rm R}(\Delta_{\rm L}-eV)f_{\rm R}(\Delta_{\rm L}-eV)-
N_{\rm R}(\Delta_{\rm L}+eV)f_{\rm R}(\Delta_{\rm L}+eV)
\right],\nonumber\\
\lim_{V\to 0}\frac{I(V,T_{\rm L}\to0)}{V}=&-\frac{2G_{\rm T}\epsilon_{\rm L}}{e}f_{\rm L}(\Delta_{\rm L})
\frac{d[N_{\rm R}(E)f_{\rm R}(E)]}{dE}
\biggr\rvert_{\Delta_{\rm L}}
\end{align}
which results to be always dissipative $I(V)>0$ when the 
product $N_{\rm R}(E)f_{\rm R}(E)$ is monotonously decreasing as it usually happens since the Fermi function is strongly decreasing with energy. 
Anyway, if the mentioned product is increasing 
around the energy $\Delta_{\rm L}$ one can still find thermoelectricity. In this case, one has an exotic example where the thermoelectric generation occur in conditions different from the one discussed in the main text. This fact demonstrates that the conditions discussed in the main text are sufficient but not necessary to have thermoelectricity. 
\subsection{Toy model II: two Dirac delta DoSs}
We consider here 
a toy model which mimic in a very simplified way the two gapped system, where the two normalized density of states are Dirac delta functions, namely $N_{\alpha}(E)=\epsilon_{\alpha}\delta(|E|-\Delta_\alpha)$, using the same notation as before for the positive constants $\epsilon_{\alpha}$. We assume $\Delta_{\rm L}\geq \Delta_{\rm R}$ with no loss of generality. The current can be computed explicitly and reads
\begin{equation}
I=\operatorname{sign}(V)\frac{G_{\rm T}\epsilon_{\rm L}\epsilon_{\rm R}}{e}
\delta(\Delta_{\rm L}-\Delta_{\rm R}-|eV|)[f_{\rm R}(\Delta_{\rm R})-f_{\rm L}(\Delta_{\rm L})].
\label{eq:IV2deltas}
\end{equation}
Consider $V>0$. The sign of the current depends on the Fermi difference in the square brackets. Since $\Delta_{\rm L}\geq \Delta_{\rm R}$, the current is always positive for $T_{\rm L}\leq T_{\rm R}$. Namely, there is no thermoelectricity if the electrode with the lower gap is the hottest, similar to what happen for a junction between two superconductors. On the contrary, for $T_{\rm L}> T_{\rm R}$ the current is negative for $V>0$ when $f_{\rm R}(\Delta_{\rm R})-f_{\rm L}(\Delta_{\rm L})<0$, which simply results in the relation 
\begin{equation}
\frac{\Delta_{\rm R}}{T_{\rm R}}>\frac{\Delta_{\rm L}}{T_{\rm L}}.
\label{eq:constraint}
\end{equation}
Namely, the thermoelectricity for this toy model arises when the temperature difference is larger than a threshold value
\begin{equation}
\Delta T=T_{\rm L}-T_{\rm R}>T_{\rm R}\frac{\Delta_{\rm L}-\Delta_{\rm R}}{\Delta_{\rm R}}.
\end{equation}
For the discussed toy model, it is easy to compute other thermodynamical quantities, such as the heat current from the hot electrode ($\dot  Q_{\rm L}$), the thermoelectric power ($\dot W=-IV$) and the thermodynamic efficiency $\eta=\dot W/\dot Q_{\rm L}$. 
In particular,
\begin{align}
\dot Q_{\rm L}&=\frac{G_{\rm T}\epsilon_{\rm L}\epsilon_{\rm R}}{e^2}\delta(\Delta_{\rm L}-\Delta_{\rm R}-|eV|)\Delta_{\rm L}[f_{\rm L}(\Delta_{\rm L})-f_{\rm R}(\Delta_{\rm R})]\nonumber\\
\dot W&=\frac{G_{\rm T}\epsilon_{\rm L}\epsilon_{\rm R}}{e^2}\delta(\Delta_{\rm L}-\Delta_{\rm R}-|eV|)(\Delta_{\rm L}-\Delta_{\rm R})[f_{\rm L}(\Delta_{\rm L})-f_{\rm R}(\Delta_{\rm R})],
\label{eq:WandQ2delta}
\end{align}
and the efficiency is 
\begin{equation}
\eta=\frac{\dot W}{\dot Q_{\rm L}}=\frac{\Delta_{\rm L}-\Delta_{\rm R}}{\Delta_{\rm L}}=1-\frac{\Delta_{\rm R}}{\Delta_{\rm L}}
\label{eq:eff2delta}
\end{equation}
Due to the constrain for thermoelectricity of Eq.~\ref{eq:constraint}, the efficiency is smaller than Carnot's limit $\eta_{\rm C}=1-T_{\rm R}/T_{\rm L}$, in accordance with the second law of thermodynamics.
\section{Voltage dependence thermoelectric power and efficiency in a SIS junction}
In the main text, we discuss the thermoelectric figures of merit in a junction between two different superconductors (SIS junction). For a given set of parameters, i.e., the temperature of the two electrodes $T_{\rm L}$, $T_{\rm R}$ and the symmetry parameter $r$, both the thermoelectric power $\dot W=-IV$ and the efficiency $\eta=\dot W/\dot Q_{\rm L}$ (we assume $T_{\rm L}>T_{\rm R}$) are function of the voltage bias $V$. In the plots of Fig.2 of the main text, we choose to evaluate these quantities at the matching peak condition, namely for $V=|\Delta_{\rm L}(T_{\rm L})-\Delta_{\rm R}(T_{\rm R})|/e$. In fact, the matching peak condition is the optimal voltage bias both for $\dot W$ and $\eta$, as we show here. Figure~\ref{FigS3} displays the voltage evolution of $\dot W$ (black curve) and $\eta$ (red curve) for the set of parameters chosen for the 
light blue curve in Fig.1 a-b of the main text. As discussed above, both $\dot W$ and $\eta$ are even in the voltage bias. By definition, $\dot W$ (and hence $\eta$) is zero at $V=0$. Moreover, $\dot W=0$ at the Seebeck voltage value $V_{\rm S}$ (where $I(V_{\rm S})=0$) and is negative at larger values (dissipative behavior). As a consequence, there is an optimal bias $V_{\rm opt}<|V_{\rm S}|$ both for $\dot W$ and $\eta$. In our system, the optimal value is the same for both the quantities and it is equal to the matching peak condition $V_{\rm opt}=V_{\rm peak}$, as shown in the plot. In this respect, the nonlinear thermoelectricity discussed here is quite different from the standard linear thermoelectricity where typically there is a trade-off in terms of the voltage bias between the power and the thermodynamic efficiency~\cite{Benenti2017}.
\begin{figure}[tp]
	\begin{centering}
		\includegraphics[width=0.6\textwidth]{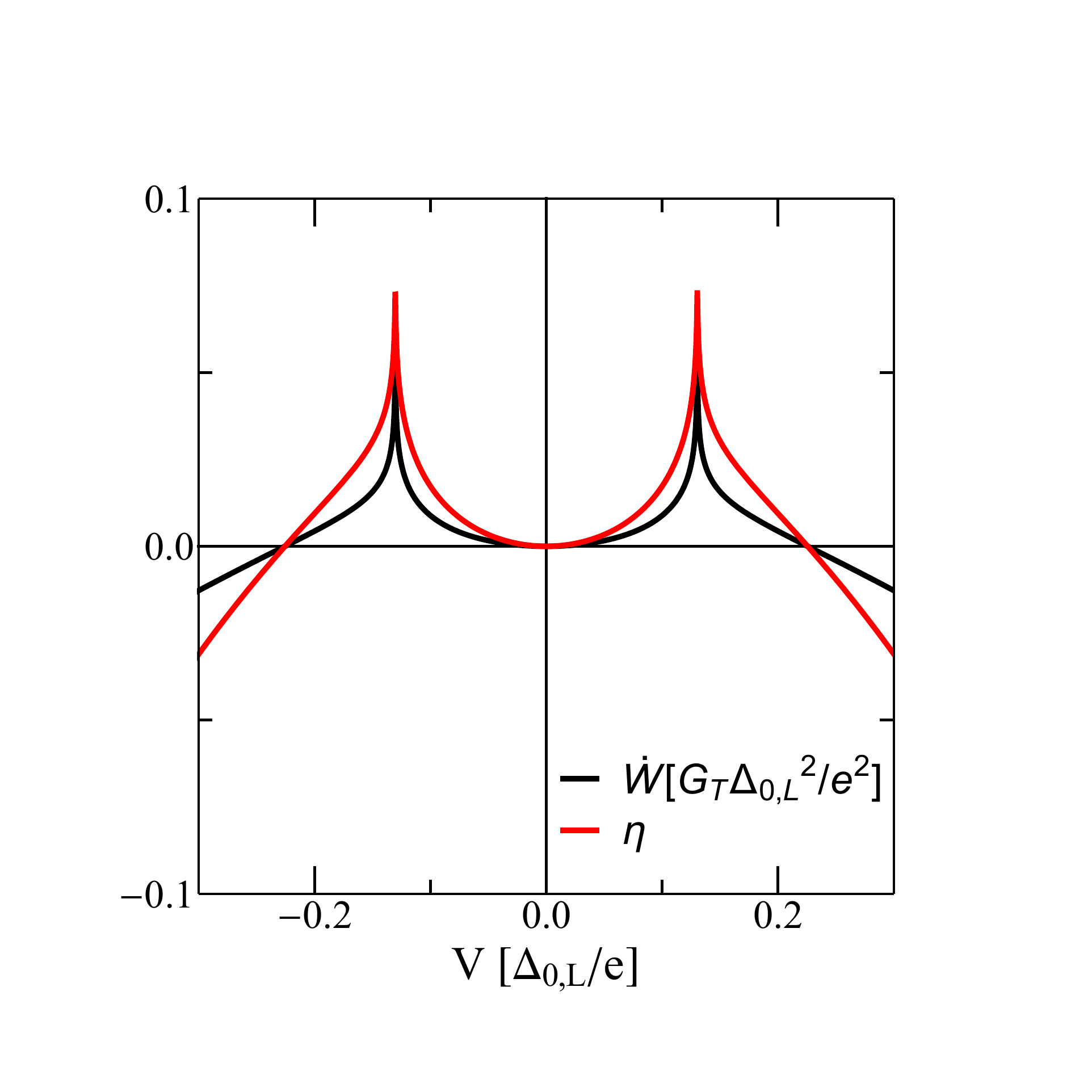}
		\caption{(color online). Voltage evolution of the thermoelectric power $\dot W=-IV$ (black) and the efficiency $\eta$ (red). Parameters are $T_{\rm L}=0.85 T_{\rm c,L}$, $T_{\rm R}=0.01 T_{\rm c,L}$, and $r=0.5$.}
		\label{FigS1}
	\end{centering}
\end{figure}
\section{Applications and stability analysis}
\begin{figure}[tp]
	\begin{centering}
		\includegraphics[width=0.8\textwidth]{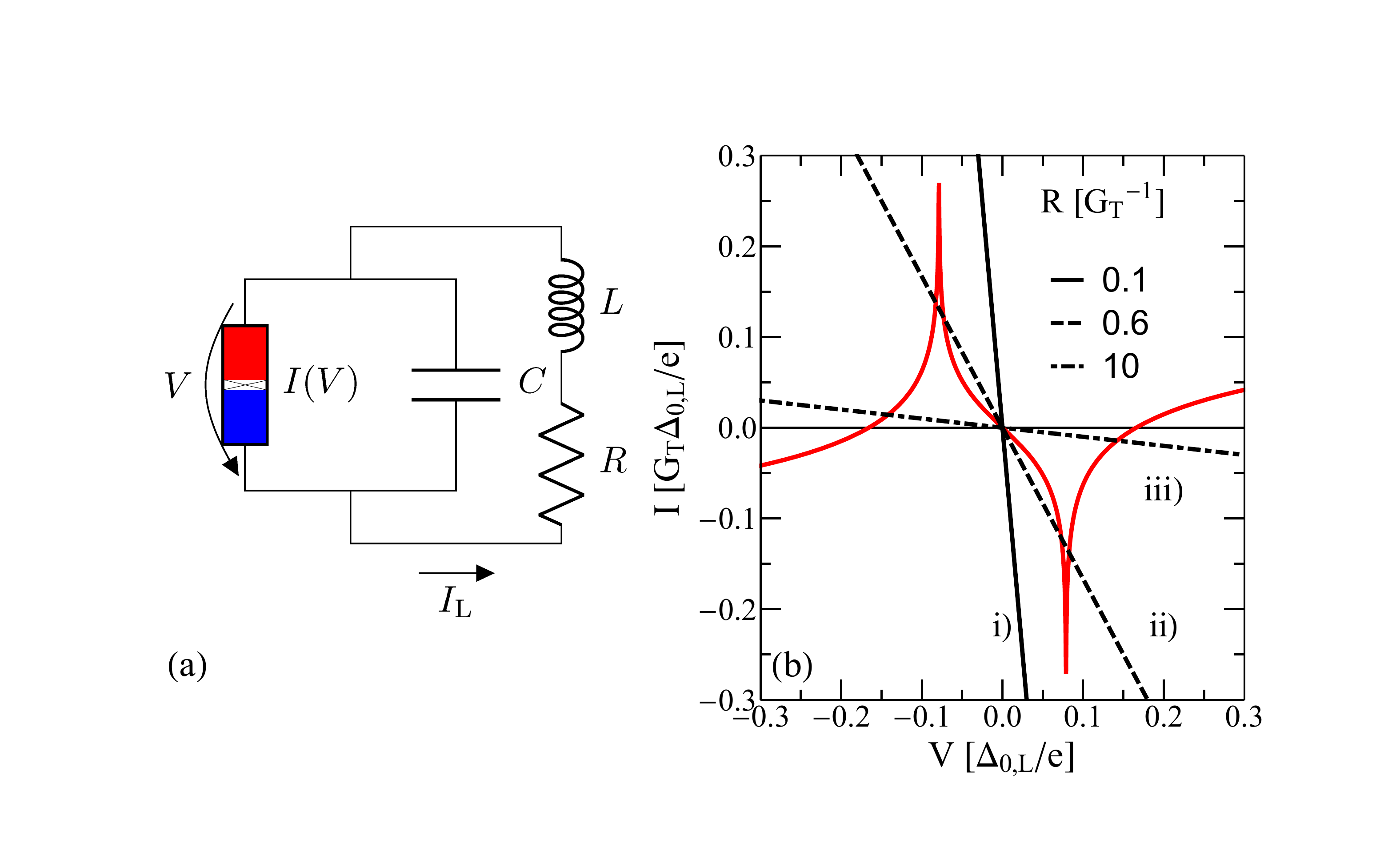}
		\caption{(color online). \textbf{(a)} Circuital scheme: the junction is a nonlinear element with characteristic $I(V)$ and capacitance $C$, connected to a generic RL circuit. \textbf{(b)} Graphical solution of Eq.~\ref{eq:implicit}. The solutions are given by the crossings of the $I(V)$ characteristic (red curve) and the load line $-V/R$ (black lines). Depending on the slope of the load line, we can have up to 5 different solutions. Below, we discuss case i) (1 solution) and case iii) (3 solutions) for the relaxation oscillator and the thermoelectric memory applications, respectively.}
		\label{FigS2}
	\end{centering}
\end{figure}
We sketch two intriguing applications of thermoelectricity in SIS junctions: i) a thermoelectric switch/memory, which relies on the existence of a multi-valued Seebeck voltage $V_{\rm S}$ at a given thermal gradient, ii) a relaxation oscillator~\cite{horowitz2015art}, based on the ANC in the $I(V)$ characteristic. To discuss both the applications, we consider the junction's dynamics in the circuit displayed in Fig.~\ref{FigS2}a. The junction is modeled as a nonlinear element of characteristic $I(V)$ and capacitance $C$, in parallel with the series of $L$ and $R$, which are the inductance and the resistance of the external circuit connected to the junction, respectively. The nonlinear dynamical system yields
\begin{equation}
\begin{cases}
I_{\rm L}=C \dot V+I(V) \\
V=-L \dot I_{\rm L}-RI_{\rm L},
\end{cases}
\label{sys:circuit}
\end{equation}
where $I_{\rm L}$ is the total current flowing in the circuit and the dot denotes the first derivative with respect to time $t$. The stationary points are obtained by setting $\dot I_{\rm L}=\dot V=0$ in Eqs.~\ref{sys:circuit}. Hence, the first of Eqs.~\ref{sys:circuit} requires $I_{\rm L}=I(V)$, which inserted in the second of Eqs.~\ref{sys:circuit} produces the implicit equation:
\begin{equation}
I(V,T_{\rm L},T_{\rm R})+\frac{V}{R}=0.
\label{eq:implicit}
\end{equation}
 Since $I(V,T_{\rm L},T_{\rm R})=-I(-V,T_{\rm L},T_{\rm R})$, the equation has an odd number of solutions. and $V=0$ is a solution of Eq.~\ref{eq:implicit} irrespectively of $R$. Moreover, in the absence of a thermoelectric effect, $IV\geq 0$ and $V=0$ is the unique solution of Eq.~\ref{eq:implicit}. In the presence thermoelectricity, we have $IV<0$ for some voltage biases and Eq.~\ref{eq:implicit} can have additional solutions.

In particular, the set of solutions of Eq.~\ref{eq:implicit} are geometrically given by the crossings of the load line $I=-V/R$ and the $I(V)$ characteristic of the junction. Figure~\ref{FigS2}b gives the graphical solution of Eq.~\ref{eq:implicit} for a set of parameters which displays thermoelectricity, i.e., $T_{\rm L}=0.7T_{\rm c,L}$, $T_{\rm R}=0.01T_{\rm c,L}$ and $r=0.75$. The number of solutions is obtained by counting the crossings and can be classified as follows
\begin{equation}
\begin{cases}
{\rm i)}\  R<|V_{\rm peak}/I(V_{\rm peak})|\quad 1\mathrm{\ solution\ (solid\ line)} \\
{\rm ii)}\ |G_0^{-1}|>R>|V_{\rm peak}/I(V_{\rm peak})| \quad 5\mathrm{\ solutions\ (dashed\ line)}\\
{\rm iii)}\ R>|G_0^{-1}|\quad 3\mathrm{\ solutions\ (dot\ dashed\ line)} \\
\end{cases}
\label{sys:numbersolutions}
\end{equation}

\begin{figure}[tp]
	\begin{centering}
		\includegraphics[width=\textwidth]{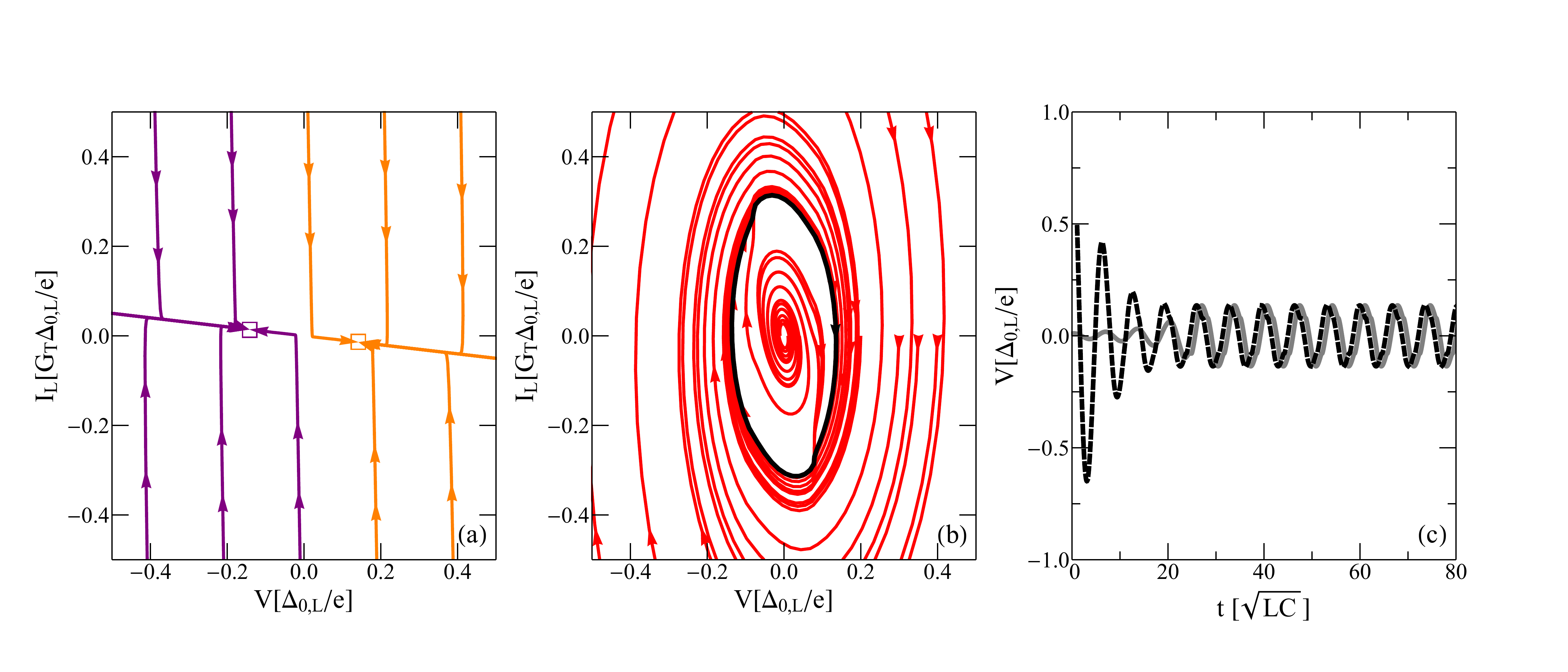}
		\caption{(color online). \textbf{(a)}  Phase portrait for $R=10 G_{\rm T}^{-1}$. All the trajectories ends up either in $(V_+,I(V_+))$ (orange-edge rectangle) or $(V_-,I(V_-))$ (purple-edge rectangle), where $V_\pm\approx\pm V_{\rm S}$. \textbf{(b)} Phase portrait for $R=0.1 G_{\rm T}^{-1}$. All the trajectories ends up in a limit cycle (black curve). \textbf{(c)} Voltage dynamics for specific initial conditions. The steady-state evolution is periodic, as expected from the phase portrait shown in panel b). Parameters are: $G_{\rm T}^{-1}=100\Omega$, $L=100$ pH,  $C=50$ fF, $T_{\rm L}=0.7T_{\rm c,L}$, $T_{\rm R}=0.01T_{\rm c,L}$, and $r=0.75$.}
		\label{FigS3}
	\end{centering}
\end{figure}

As we will see below, we are mainly interested in case i) and case iii) for the relaxation oscillator and the memory application, respectively. For simplicity, we discuss the stability of the stationary points with a linearization procedure~\cite{strogatz2014nonlinear}. In particular, upon linearizing the system of Eqs.~\ref{sys:circuit} around the generic solution $V=\tilde V$ of Eq.~\ref{eq:implicit}, we get
\begin{equation}
\frac{d}{dt}
\begin{pmatrix}
v   \\   
i_{\rm L } 
\end{pmatrix}
=A
\begin{pmatrix}
v   \\   
i_{\rm L} 
\end{pmatrix}=
\begin{pmatrix}
-G(\tilde V)/C   & 1/C \\   
-1/L  & -R/L 
\end{pmatrix}
\begin{pmatrix}
v   \\   
i_{\rm L} 
\end{pmatrix}
\label{sys:linearized}
\end{equation}
where $v=V-\tilde V$, $i_{\rm L} =I_{\rm L} -I(\tilde V)$ and  $G(\tilde V)= dI/dV|_{V=\tilde V}$. Clearly, $G(\tilde V=0)=G_0$. The solution of the linearized system can be obtained by computing the eigenvalues $\lambda_{\pm}$ of the 2x2 matrix $A$ of Eqs.~\ref{sys:linearized}. They reads:
\begin{equation}
\lambda_{\pm}=\frac{\Sigma}{2}\pm\sqrt{\left(\frac{\Sigma}{2}\right)^2-D}=0,
\label{eq:eigenvalues}
\end{equation}
where $\Sigma={\rm Tr} [A]=-G(\tilde V)/C-R/L $ and $D={\rm Det} [A]=(G(\tilde V)R+1)/(LC)$ are the trace and the determinant of the $A$ matrix, respectively. The stationary point is stable if the real parts of both the solutions $\lambda_{\pm}$ are negative~\cite{strogatz2014nonlinear}.
For $D<0$, we observe that the two solutions of Eq.~\ref{eq:eigenvalues} are reals and they have opposite signs, i.e., the stationary point is unstable. Conversely, for $D>0$, the signs of the real parts of $\lambda_\pm$ are equal to the sign of $\Sigma$. As a consequence, one gets a stable solution when $\Sigma<0$ and $D>0$. Moreover, when $D>\Sigma^2/4$, the eigenvalues get also an imaginary component which gives the frequency of oscillation of the overdamped evolution when the system evolves toward the stationary solution.

With respect to the system of Eqs.~\ref{sys:linearized}, we note that $G(\tilde V)>0$ implies $\Sigma<0$ and $D>0$, hence each stationary point with positive conductance is stable, corresponding to a dissipative behavior. Conversely, the case $G(\tilde V)=-|G(\tilde V)|<0$ gives a richer phenomenology. Since $\Sigma< 0$ gives $R>|G(\tilde V)|L/C$ and $D>0$ implies $R<|G(\tilde V)|^{-1}$, it is possible to have values of $R$ which give a stable solution only when $|G(\tilde V)|\sqrt{L/C}<1$. 

We are now ready to discuss two different regimes. 

\textbf{Thermoelectric switch/memory-case iii).}
For large values of the load, Eq.~\ref{eq:implicit} has three solutions $V=0,\tilde V_{\pm}$. The solution $V=0$ is unstable since $G(\tilde V=0)=-|G_0|<0$ and the condition $R<|G(\tilde V)|^{-1}=|G_0|^{-1}$ is violated, as discussed in the classification of Eqs.~\ref{eq:eigenvalues}. Conversely, the two solutions $V= \tilde V_{\pm}$ are stable since $G(\tilde V_{\pm})>0$. Therefore, each trajectory in the phase portrait evolves either toward $\tilde V_+$ or $\tilde V_-$. Which one of the two solutions, depends on the initial conditions. This behavior is displayed in the phase portrait of Fig.~\ref{FigS3}a, obtained for $R=10 G_{\rm T}^{-1}$. As discussed in the main text in a simplified circuit, the combination of the nonlinear temperature bias and the voltage associated to the load of the external circuit leads to a spontaneous breaking of EH symmetry. The system acts as a thermal switch, where the output signal is provided if the temperature gradient produces an ANC. Moreover, the voltage bi-stability can be exploited to design a thermoelectric memory. When the system is polarized either in $V_\pm$, the signal persists even after removing the external voltage. However this kind of memory is volatile, i.e., the system relaxes to $I_{\rm L}=V=0$ when the thermal gradient is removed.

\textbf{Thermoelectric oscillator-case i).}
When the load is very small, there is a unique solution of Eq.~\ref{eq:implicit}, i.e., the trivial solution $V=0$, where $G(\tilde V=0)=-|G_0|<0$. In the numerical calculation, we consider a set of realistic parameters which satisfy the inequality $|G_0|\sqrt{L/C}>1$. As a consequence, the zero-current state with $V=0$ is unstable, giving rise potentially to an oscillatory behavior. Figure~\ref{FigS3}b displays the phase portrait for $R=0.1 G_{\rm T}^{-1}$. The plot looks quite different compared to Fig.~\ref{FigS3}a. In particular, all the trajectories collapse on a close curve in the phase plane, known as the \textit{limit cycle} (black curve). Hence, the system displays self-sustained oscillations after a transient dynamics, as better visualized in Fig.~\ref{FigS3}c for some initial conditions. Namely, the circuit acts as a \textit{thermoelectric relaxation oscillator}, i.e., a system whose electrical oscillations are produced by the presence of a thermal gradient. The period and the shape of the steady-state oscillations depend strongly on the features of the $I(V)$ function, as we discuss now. 

\subsection{Existence and uniqueness of the limit cycle.}
For simplicity, we focus on the limit $R\rightarrow 0$ and we write a second order differential equation for $V$ by combining the two equations of the system of Eqs.~\ref{sys:circuit} 
\begin{equation}
\ddot V+\sqrt\frac{L}{C}G(V)\dot V+V=0,
\label{eq:2ndV}
\end{equation}
where the dot gives the time derivative with respect to the dimensionless time $\tau=t/\sqrt{LC}$. Note that $G(V)=dI(V)/dV$ is an even function of $V$ since $I(V)$ is an odd function. As a consequence, Eq.~\ref{eq:2ndV} is of the Li{\'e}nard type $\ddot x+f(x)\dot x+g(x)=0$, where $f(x)$ is an even function and $g(x)$ is an odd function. 

For $g(x)=x$, as in our case, a generalization of the Levinson-Smith theorem states that a dynamical system of the Li{\'e}nard type displays a unique stable cycle limit under the following assumptions~\cite{Sansone1949,Sabatini}:
\begin{itemize}
	\item exists $x_0>0$ such that $f(x)<0$ for $|x|<x_0$
    \item either $f(x)>0$ in $(x_0,\infty)$ or $f(x)>0$ in $(-\infty,-x_0)$ 
    \item exists $x_1>0$ such that $F(x_1)=F(-x_1)=0$ 
    \item $\lim_{x\rightarrow\pm \infty} F(x)=\pm \infty$
\end{itemize}
where $F(x)$ is the primitive of $f(x)$. 

In our case $f(x)=G(V)$, $F(x)=I(V)$.  Note that the last condition is always satisfied, since $I(V)\sim G_{\rm T}V$ at large $V$. In the presence of a thermoelectric effect, we have that $G(V)<0$ for $|V|<V_{\rm peak}$ and positive otherwise. As a consequence, all the remaining assumptions are satisfied with $x_0=V_{\rm peak}$ and $x_1=|V_{\rm S}|$.

In general, the properties of the steady state oscillation, i.e., the period and the shape of the signal, depend on the details of the nonlinear term $G(V)$. This can be understood by looking at standard examples of negative resistance oscillators, such as the Van Der Pol oscillator or the piecewise linear oscillator~\cite{stoker1950nonlinear,strogatz2014nonlinear}.
Two limits are typically discussed:

\textbf{Small non-linearity.} As a first approximation, one can solve Eq.~\ref{eq:2ndV} by neglecting the second term when the non-linear term $|G(V)\sqrt{L/C}|$ is small. The solution of the linear equation is a sinusoidal oscillation $V(t)=\mathcal A\sin(\omega t)$ with characteristic frequency $\nu=\omega/(2\pi)=1/(2\pi\sqrt{LC})$. Therefore, the periodic signal has a frequency $\nu=1/\tau_{\rm per}$ in the range 10-100 GHz for realistic parameter values. Note that, unlike a proper linear system, the amplitude of the oscillations $\mathcal A$ is fixed by the non-linearity of the problem, since the limit cycle is unique.
 
\textbf{Large non-linearity.} The steady-state is characterized by a slow-fast dynamics. The shape of the signal depends strongly on the non-linear term and the period can be computed, in simple cases, through a perturbative approach. Typically the frequency scales with the strength of the non-linear term and hence it is reduced with respect to the almost-linear situation. This is well known and described for standard systems like the Van Der Pol oscillator~\cite{stoker1950nonlinear}.